\documentclass[sn-mathphys,Numbered]{sn-jnl}% Math and Physical Sciences Reference Style
\usepackage{graphicx}%
\usepackage{float}%
\usepackage{multirow}%
\usepackage{amsmath,amssymb,amsfonts}%
\usepackage{amsthm}%
\usepackage{mathrsfs}%
\usepackage[title]{appendix}%
\usepackage{xcolor}%
\usepackage{textcomp}%
\usepackage{manyfoot}%
\usepackage{booktabs}%
\usepackage{algorithm}%
\usepackage{algorithmicx}%
\usepackage[noend]{algpseudocode}%
\usepackage{listings}%
\usepackage{bm}%
\usepackage{mathtools}%
\usepackage{scalerel}%
\newcommand{\myMatrix}[1]{\bf{\mathit{#1}}}
\newcommand{\m}{\scalebox{0.5}[1.0]{\( - \)}}
\usepackage{booktabs}
\usepackage{import}

\theoremstyle{thmstyleone}%
\theoremstyle{thmstyletwo}%

\theoremstyle{thmstylethree}%

\begin{document}

%\title[Article Title]{High-Performance Ising Models for Wafer-Scale Engines}
\title[Article Title]{Record Acceleration of the Two-Dimensional Ising Model Using High-Performance Wafer Scale Engine}


\author[1]{\fnm{Dirk} \sur{Van Essendelft}}\email{dirk.vanessendelft@netl.doe.gov}

\author[2]{\fnm{Hayl }\sur{Almolyki}}

\author[3]{\fnm{Wei}\sur{Shi}}\email{Wei.Shi@netl.doe.gov}

\author[4]{\fnm{Terry}\sur{Jordan}}
\author[5]{\fnm{Mei-Yu}\sur{Wang}}
\author[6]{\fnm{Wissam A.}\sur{Saidi}}\email{Wissam.Saidi@netl.doe.gov}

\affil[1,3,4]{\orgdiv{Advanced Computing and Artificial Intelligence}, \orgname{The National Energy Technology Laboratory}, \orgaddress{\street{3610 Collins Ferry Rd}, \city{Morgantown}, \postcode{26505}, \state{WV}, \country{USA}}}
\affil[2,6]{\orgdiv{Computational Materials Engineering}, \orgname{The National Energy Technology Laboratory}, \orgaddress{\street{626 Cochrans Mill Rd}, \city{Pittsburgh}, \postcode{15236}, \state{PA}, \country{USA}}}

\affil[5]{\orgdiv{HPC AI and Big Data Group}, \orgname{Pittsburgh Supercomputing Center, Carnegie Mellon University}, \orgaddress{\street{300 S. Craig St.}, \city{Pittsburgh}, \postcode{15213}, \state{PA}, \country{USA}}}

\abstract{
The versatility and wide-ranging applicability of the Ising model, originally introduced to study phase transitions in magnetic materials, have made it a cornerstone in statistical physics and a valuable tool for evaluating the performance of emerging computer hardware. Here, we present a novel implementation of the two-dimensional Ising model on a Cerebras Wafer-Scale Engine (WSE) – a revolutionary processor that is opening new frontiers in computing. In our deployment of the checkerboard algorithm, we optimized the Ising model to take advantage of the unique WSE architecture. Specifically, we employed a compressed bit representation storing 16 spins on each $int16$ word, and efficiently distributed the spins over the processing units enabling  seamless weak scaling and limiting communications to only immediate neighboring units. Our implementation can handle up to 754 simulations in parallel, achieving an aggregate of over 61.8 trillion flip attempts per second for Ising models with up to 200 million spins. This represents a gain of up to 148 times over previously reported single-device with a highly optimized implementation on NVIDIA V100 and up to 88 times in productivity compared to NVIDIA H100. Our findings highlight the significant potential of the WSE in scientific computing, particularly in the field of materials modeling.
}

\keywords{Wafer Scale Engine, High Performance Computing, Ising model}


\maketitle

\section{Introduction}\label{sec1}
Advanced materials are a crucial enabling technology for humankind to remain competitive in addressing challenges related to energy, health, and nanotechnology. While materials modeling based on atomistic principles has made significant strides in the past decade in designing novel materials for different applications, there are still considerable challenges owing to the computational cost of applying these models, particularly for large and complex systems.  These challenges call for the development of new methodologies as well as in advancement of software and hardware technologies. The Wafer Scale-engine (WSE), the world’s largest accelerator chip in existence, represents a significant hardware breakthrough in the field of computing and integrated circuit design though its impact on scientific computing is yet to be fully assessed.\cite{Cerebras_story, Cerebras_White_paper1, Cerebras_White_paper2}

\begin{figure}[h!]
    \centering
  \includegraphics[width=\textwidth]{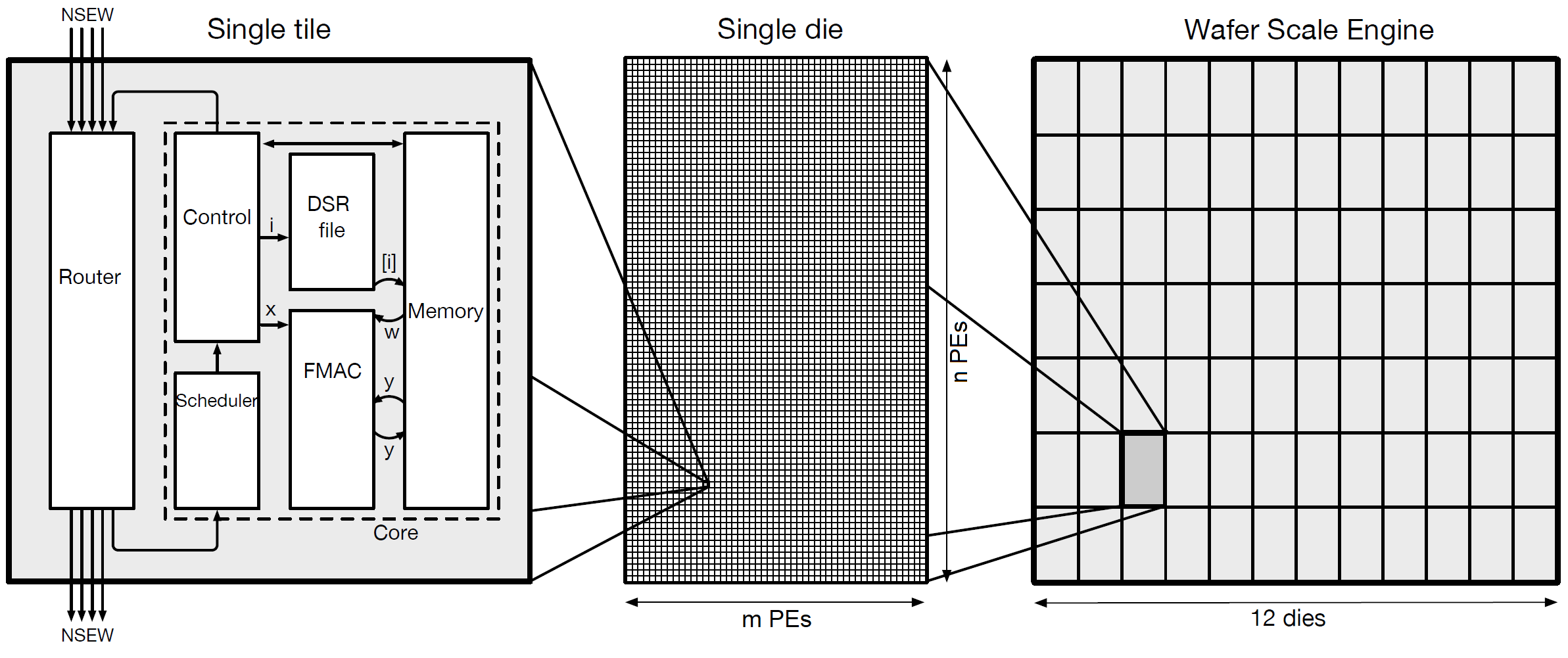}
    \caption{CS-2 Wafer scale engine. (rightmost) A single Wafer Scale Engine is a single processor spanning the largest possible square that can be patterned on a 300mm wafer. Each processing engine (PE) is a Turing-complete, independently programmable computer consisting of a controller, Arithmetic Logic Unit (ALU), 48KB of Static Random-Access Memory (SRAM), and a router that can communicate with nearest-neighbor PEs. All main memory in WSE is SRAM and accessible at L1 cache rates (128b of read and 64b of write on each cycle), which is matched to the ALU processing rates. (middle) Each processor is a collection of dies arranged in a 2D fashion that are then further subdivided into a grid of Processing Elements (PEs). (leftmost) One die hosts thousands of PEs (computational cores, memory and routers). There is no logical discontinuity between adjacent dies and there is no additional bandwidth penalty for crossing the die-die barrier.}  
    \label{fig:WSE_diagram}
\end{figure}

As schematically shown in  Figure~\ref{fig:WSE_diagram}, the WSE is a massively-parallel, data-flow architecture consisting of a grid of nearly a million, identical Processing-Elements (PEs) within nearly a 215 mm $\times$ 215 mm wafer, without interposers and chiplets arranged on a two-dimensional (2D) grid across the wafer surface. Each PE is a Turing-complete, independently programmable computer consisting of a controller, Arithmetic Logic Unit (ALU), 48KB of Static Random-Access Memory (SRAM), and a router that can communicate with nearest-neighbor PEs. We conducted our work on a CS-2 and configured various fabric sizes ranging from $130 \times 133$ up to $756 \times 993$.

On traditional compute hardware latency is unfavorably high (hundreds of system clock cycles from processor to memory and many thousands of cycles between processors on a network) and memory/network bandwidths are several orders of magnitude lower than L1 cache rates. By contrast, all main memory in the WSE is SRAM and accessible at L1 cache rates (128b of read and 64b of write on each cycle), which is matched to the ALU processing rates and the on-chip network bandwidth.  On a system level, PEs on the WSE can coordinate to achieve a larger task via the massively-parallel, data-flow architecture.  The WSE is designed to be able to trigger computations based on the arrival of data or control signals.  Each PE supports fused operations where one operands data (64b) can come from the network-on-chip router and one can come from local memory.  Data can be simultaneously sent, received, and processed via micro-thread scheduling.  Synchronous machine operation is achieved by unblocking/activation of tasks at micro-thread processing conclusion. \cite{rocki2020Fast}

The WSE, Field-equation, Application-programming-interface (WFA) was developed as an easy-to-use interface to programming the WSE. Specifically, the WFA is designed to support and accelerate models that can be distributed on a structured grid with hexahedral connectivity including computational fluid dynamics, structural mechanics, and subsurface modeling.\cite{wfa_program}
Recent benchmarks in computational fluid dynamics showed two to three orders of magnitude speedup relative to a typical CPU cluster.\cite{rocki2020Fast, woo2022disruptive}

The Ising model\cite{Ising1925BeitragZT} initially devised to study phase transitions in magnetic materials stands as one of the most fundamental models in statistical physics to investigate the universal behavior of critical phenomena. The Hamiltonian of the Ising system is defined as, $H=-J\sum_{\langle i, j\rangle} \sigma_{i} \sigma_{j}$, where $J$ is the coupling strength of interaction between nearest neighbors $\langle i, j\rangle$ spins $\sigma_{i}$ and $\sigma_{j}$. For example, in 2D systems, each spin interacts with $4$ neighbors while in 3D each spin interacts with 6 neighbors.  $\sigma_{i}$ can assume values $+1$ or $-1$ and thus the $\sigma_{i}$ $\sigma_{j}$ interaction can assume a value of $+1$ or $-1$ when spins are parallel or antiparallel, respectively.  For ferromagnetic coupling with $J>0$, the total system energy is minimized if all spins are aligned or parallel.  The 1D Ising model has no phase transition to a ferromagnetic state at any finite temperature, while in higher dimensions, it undergoes a thermally-driven order-to-disorder phase transition.\cite{Landau2014book} The critical temperature $T_C$ at which the phase transition occurs has been analytically determined for 2D Ising model by Onsager.\cite{Onsager1944PhysRev} 
For higher dimensions, the critical temperature is not known analytically but can be determined numerically based on Monte Carlo simulations.\cite{Landau2014book}

The Monte Carlo (MC) approach is a widely employed method for solving complex systems like the Ising model in statistical physics, which relies heavily on the Metropolis probability criterion and random number generation.\cite{Landau2014book} In the MC approach, an initial spin is randomly selected, and an attempt is made to flip its sign that is accepted based on Metropolis probability $e^{-\beta \Delta E}$ where $\beta=1/k_{B} T$,  $k_B$ is the Boltzmann constant, $T$  is temperature, and $\Delta E$ is the energy change in the system due to a spin flip. If the proposed spin reduces the energy or satisfies the probabilistic Metropolis decision then the move is accepted and a new configuration is generated, otherwise the system retains the old configuration. Metropolis probabilistic decision is based on a random number $ran()$ in $[0,1)$ and the flip is accepted if $ran() < e^{-\beta \Delta E }$.  
%By generating random numbers and using the Metropolis algorithm to probabilistically accept or reject spin flips, MC simulations of the Ising model provide a statistical representation of the system's behavior at a given temperature.  

Due to their locality, the sequential single-site MC approach suffers from the problem of critical slowing down near $T_C$, as it becomes increasingly difficult to flip a spin at random that is likely to be coupled to neighboring spins pointing in the same direction.This leads to the divergence of the relaxation times near $T_C$. The critical slowing down was mitigated by cluster algorithms such as Swendsen-Wang\cite{Swendsen87PRL} and Wolff\cite{Wolff89PRL}, where an attempt is made to flip a cluster of spins based on Metropolis probability rather than a single spin. Cluster algorithms significantly reduce the number of updates needed to reach equilibrium and provide more accurate estimates of thermodynamic properties, especially in the vicinity of phase transitions. 

The Ising model's combination of simplicity and complexity, along with its suitability for parallelization and algorithm optimization, makes it a valuable tool for testing and benchmarking new hardware in the realm of computational physics and high-performance computing (HPC).   The Ising model has been ported on different computing hardwares including GPUs \cite{Block12EPJ,Komura12JCP,Romero2020CPC}, Tensor Processing Units (TPU) \cite{Yang19GPUarxiv}, and Field Programmable Gate Arrays (FPGAs) \cite{Ortega-Zamorano16IEEE,LinWZ13JCP,Gilman13proceeding}.  The comparison between the speedups obtained on different architectures is convoluted due to different algorithms employed to solve the Ising model, different metrics involved in quantifying the performance, and whether the obtained metrics are dated considering new architectures.  Typically, the number of attempted spin flips per nanosecond (flip/ns) is used as a metric where high flips/ns is indicative of a better performance. The 2019 TPU v3 implementation on single core reported 12.8 flips/ns and over multiple cores reported 11.43 flips/ns/core (e.g., 366 flips/ns over 32 cores).\cite{Yang19GPUarxiv} In comparison, the 2020 implementation on single Tesla V100 GPU reported 66.954 flips/ns using a basic CUDA C implementation.\cite{Romero2020CPC} Further, this study optimized the code by using a four~bit representation per spin, and reported an impressive rate of $459$~flips/ns on a single Tesla V100 GPU.  The multi-GPU implementation of up to 16 GPUs achieved 7381 flips/ns. A FPGA implementation achieved a speed-up factor of approximately 104 times in comparison to a standard CPU simulation and 614 flips/ns for a lattice of $1024\times1024$ spins in 2016.\cite{Ortega-Zamorano16IEEE} 

Herein, we implement the 2D Ising model on the WSE and optimize the code using a novel approach to take advantage of its unique hardware architecture. We show that the WSE can simultaneously handle up to 754 simulations in parallel and achieve up to 148 times speedup over previously reported single-device on NVIDIA V100 and up to 88 times in productivity compared to a highly optimized code on NVIDIA H100. Additionally, our deployment achieved perfect weak scaling out to the maximum wafer extents owing to the optimum implementation that requires only communications between nearest-neighbor processing units. 

\section{Implementation of the 2D Ising model on the WSE}\label{wse_implement}

\begin{figure}[h!]
    \centering
  \includegraphics[width=\textwidth]{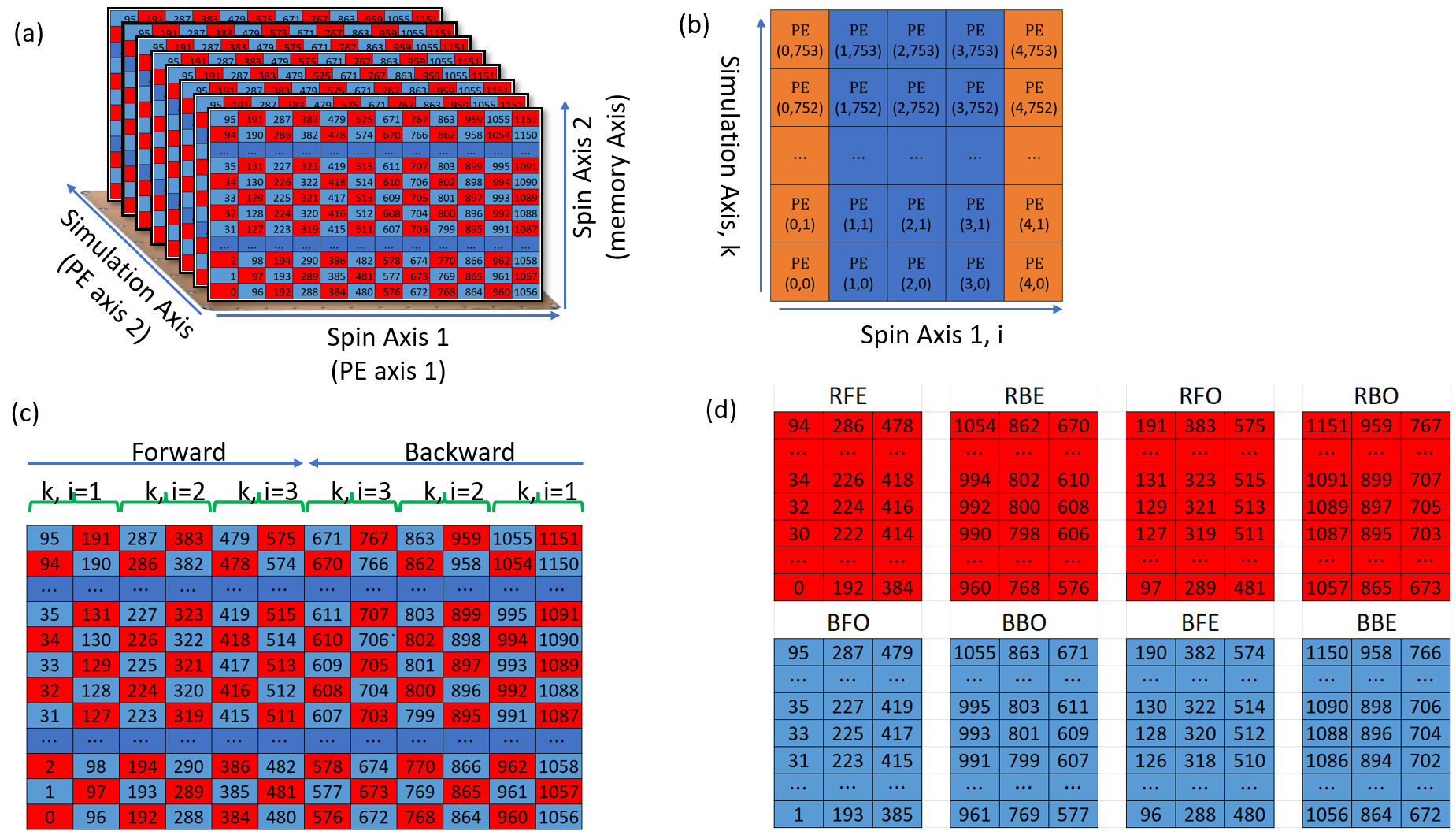}
    \caption{ Conceptual layout of the $12 \times 96$ 2D Ising model on WSE hardware. (a) The WSE is used to solve multiple instances of the 2D Ising model, where each instance is solved independently on a single column of the WSE. Note that the real WSE-2 architecture contains nearly 1 million PEs distributed over 756 columns.  We devote one axis of PE’s to one of the axis of spins in the 2D Ising model, the second axis is devoted to simulations in parallel.  The second spin axis is held in the memory of each PE.  (b) The conceptual layout within the WFA with a single fold.  Each simulation spans 5 PEs (3 workers, blue, and 2 moats, orange).  The expanded and PBC updated column vectors are spread across the spin axis in order.  The moats hold one axis of boundary condition data.  The second axis of BC data is held in the top and bottom of each vector on the workers memory space. 
    (c) Checkerboard decomposition. (d) 8-array representation on the WSE.  For instance, the RFE array refers to spin belong to red checkerboard in (a) with even index and forward order. BBO refers to spins belonging to blue checkerboard with odd index and backward ordering. The mapping between the Ising lattice and the WSE PEs are also shown in (b) where $k,i$ refers to PE $(x,y)$ coordinates.  For simplicity and clarity, we only show few spin indices from 0 to 1152.}
    \label{fig:decomposition}
\end{figure}

Naively, given the local interaction between nearest neighbors in the Ising model, it is tempting to distribute the spins such that each PE will have a single spin. While simple in concept, this arrangement results in very low memory utilization and very high parasitic cycle consumption.  A single spin will only consume one bit of the 48kB available per PE.   In addition, it takes tens of cycles of operations to set up an arithmetic operation. Operating on single spin element is the least efficient way to use the PE, and it is computationally far more efficient to use the memory to its fullest extent and operate over longer vectors to dilute instructions overhead. With this in mind, our approach for solving the Ising model utilizes only a single PE column of the 2D WSE grid. As we elaborate below, we have devised a computationally efficient approach to distribute the spins that requires minimal communication between the PE's.  Further, different columns of the PE's on the WSE can then be utilized to simultaneously solve different 2D Ising models, e.g., at different temperatures as implemented in our study. See schematic in Figure~\ref{fig:decomposition}a.

The checkerboard decomposition of the Ising model allows for efficient parallel processing as the local adjacency matrix for each spin shows dependence only on the opposite color when attempting to flip a spin with a specific color, i.e.  flipping red spins depends only on neighboring blue spins and vice versa. This is illustrated for a $12 \times 96$ lattice in Figure~\ref{fig:decomposition}. Thus, an $m \times n$ Ising lattice can be conveniently split into two equal arrays of size $mn/2$ and the spins within each half are operated on in parallel. See Supplementary information for single-spin and checkerboard algorithms. Romero et al. implemented the checkerboard algorithm on GPUs and optimized it by representing a single spin as a four-bit word to reduce memory traffic and storage per spin.\cite{Romero2020CPC} 

To leverage the hardware architecture of the WSE, our implementation casts the Ising lattice into eight arrays and compacts 16 spins into a single $int16$ word. This decomposition involves the standard color separation of the checkerboard into red (R) and blue (B), followed by further dividing each color into even (E) and odd (O) arrays based on spin index. Subsequently, we utilize domain folding by halving these four arrays and reversing the order of the right half, resulting in forward (F) and backward (B) spin orderings. We refer to this as domain folding with $N_{\rm{fold}}=1$. See Figure~\ref{fig:decomposition}d for visual representation for the case of $12 \times 96$ Ising lattice. For clarity, we use a three-letter code to denote the eight arrays. For example, "RFE" signifies "Red-Forward-Even" and "BBO" represents "Blue-Backward-Odd." This 8-array representation mandates that the dimensions of the Ising lattice $m \times n$, must be selected such that $m$ is a multiple of 4 and $n$ a multiple of 32. 

Domain folding with the forward and backward spin arrangements, ensures that updates across periodic boundaries involve communication solely with immediate neighboring PEs. Thus, this decomposition, specifically tailored to take advantage of the unique WSE architecture, enables perfect weak scaling by eliminating the need to relay messages from one end of the wafer to the other, as demonstrated later. In fact, the 2D Ising lattice domain folding can be applied any odd number $N_{\rm{fold}}$, to fine-tune the memory allocation between width and height in a 2D lattice model. For example, some of our benchmarking discussed below is done with $N_{\rm{fold}}=5$.

\begin{figure}[h!]
    \centering    \includegraphics[width=\textwidth]{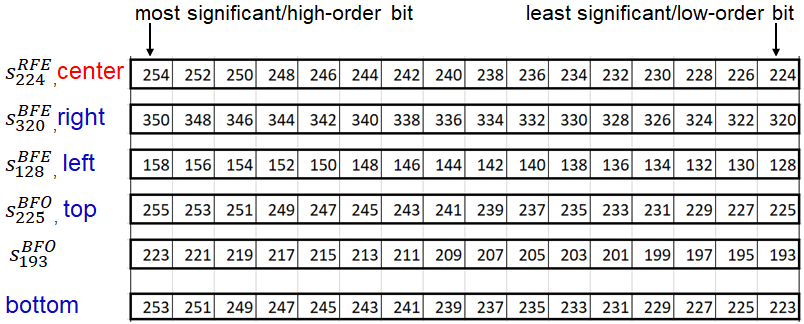}
    \caption{An illustration of the compacted spin representation around $s_{224}^{RFE}$ for the $12 \times 96$ lattice decomposition. All values displayed are $int16$ integer values that contain 16 bits representing 16 spin states in an ascending order from least to most significant bit.  
    %The spins correspond to the checker mapping in Figure~\ref{fig:decomposition}. 
    $s_{224}^{RFE}$ contains 16 bits representing spins between 224 (least significant) and 254 (most significant).  
    %All spins in $s_{224}^{RFE}$ are red.  
    The right, left, and top neighbors to all spins in $s_{224}^{RFE}$ are blue spins in $s_{320}^{BFE}$, $s_{128}^{BFE}$, and $s_{225}^{BFO}$, respectively.  The bottom spins to $s_{224}^{RFE}$ can be derived from $s_{225}^{BFO}$ and $s_{193}^{BFO}$ using logical operators as shown in Algorithm \ref{alg:pe_getNeighbors}. 
    %Note that there is no $int16$ integer in Eq \ref{eq:BFO_vect_bcu} to represent the 16-bit bottom string.
    }
\label{fig:int16_bit_neighboring}
\end{figure}

To simplify the notation, we introduce a compact representation in which a single 16-bit integer, $s$, that can hold 16 spins is tagged with the three-letter code as a superscript and the spin index in the original lattice as a subscript. The bit value for this spin index occurs at the least significant bit position in the 16-bit representation (Figure \ref{fig:int16_bit_neighboring}). For example, $s_{0}^{\scaleto{RFE}{4pt}}$ contains spins [30, 28, ... 2, 0] where spin 0 is used as the subscript as it sits at the least significant bit (rightmost) in the $int16$ word. Using this notation, a $12 \times 96$ Ising lattice can be described using eight $3\times3$ $int16$ arrays, as demonstrated for the RFE case (Figure~\ref{fig:decomposition}) by $\mathbf{S}^{\scaleto{RFE}{4pt}}$ in Eq.~\ref{eq:RFE_vect_12x96}.  See supplementary information for the $3\times3$ representations of the other arrays. 

\begin{equation}\label{eq:RFE_vect_12x96}
\mathbf{S}^{\scaleto{RFE}{4pt}} = \begin{bmatrix}
s_{64}^{\scaleto{RFE}{4pt}} & s_{256}^{\scaleto{RFE}{4pt}} & s_{448}^{\scaleto{RFE}{4pt}}\\
s_{32}^{\scaleto{RFE}{4pt}} & s_{224}^{\scaleto{RFE}{4pt}} & s_{416}^{\scaleto{RFE}{4pt}}\\
s_{0}^{\scaleto{RFE}{4pt}}  & s_{192}^{\scaleto{RFE}{4pt}} & s_{384}^{\scaleto{RFE}{4pt}}\\
\end{bmatrix}
\end{equation} 
 
$\mathbf{S}$ facilitates the identification of neighboring spins on the Ising lattice through straightforward, and importantly, computationally efficient bit operations, as illustrated in Algorithm~\ref{alg:pe_getNeighbors}.  Specifically, since spins are ordered sequentially with an increment of 2, neighboring spins located at the least significant bits imply that all spins/bits within the $int16$ word are also neighbors. We illustrate this for $\mathbf{S}^{\scaleto{RFE}{4pt}}[1,1]=s_{224}^{\scaleto{RFE}{4pt}}$. 
For this case, the nearest neighbors are determined by inspecting only the blue odd and even matrices $\mathbf{S}^{\scaleto{BFE}{4pt}}$ and $\mathbf{S}^{\scaleto{BFO}{4pt}}$ to identify integers with the least significant bit/spin associated with neighboring spins. Figure~\ref{fig:decomposition}c shows that  the right, left, top, and bottom neighboring spins of 224 are respectively 320, 128, 225, and 223.  Three of the neighbors can be found in the arrays with no need for manipulation ($\mathbf{S}^{\scaleto{BFE}{4pt}}[1,1]=s_{320}$, $\mathbf{S}^{\scaleto{BFE}{4pt}}[0,1]=s_{128}$, $\mathbf{S}^{\scaleto{BF0}{4pt}}[1,1]=s_{225}$). The bottom nearest neighbor, 223, can be obtained using common bitwise operators.  $\mathbf{S}^{\scaleto{BFO}{4pt}}[1,0]=s_{193}^{\scaleto{BFO}{4pt}}$ is right shifted by 15 bits, $\mathbf{S}^{\scaleto{BF0}{4pt}}[1,1]=s_{225}$ is left shifted by 1 bit, and the shifted integer arrays are combined with a bitwise OR (the result of this operation is shown in Figure~\ref{fig:int16_bit_neighboring}). 

Performing similar local neighborhood analyses for each of the eight arrays reveals data access patterns that can be utilized in the code implementation kernels, as summarized in the following four rules. In this notation, $i$ and $j$ represent the first and second dimensions in the compacted spin arrays (i.e., $\mathbf{S}^{\scaleto{BFO}{4pt}}[i,j]$). Also, $i,j$ are consistent with the Spin Axis directions depicted in Figure \ref{fig:decomposition}a,b.
\begin{enumerate} 
  \item 
  %%\hspace{10} %%[Rule 1.]
  RFE, RBO, BBE, BFO  are only accessed from $i$ and $i+1$ positions
  \item %%[Rule 2.]
  RBE, RFO, BFE, BBO  are only accessed from $i$ and $i-1$ positions
  \item %%[Rule 3.]
  Even arrays are only accessed from $j$ and $j+1$ positions
  \item %% [Rule 4.] 
  Odd arrays are only accessed from $j$ and $j-1$ positions
\end{enumerate}
%\end{itemize}

To use stencil operations beyond the central elements of the eight arrays, we expand $\mathbf{S}$ by one in all directions and copy data from neighbors at appropriate times during the algorithm.  This approach allows us to leverage single vector instructions to fully compute to the original extents of the unexpanded arrays at minor memory and PE costs.  The expanded and periodic boundary condition (PBC) updated array for the RFE array is shown in Eq.~\ref{eq:RFE_vect_bcu}. See Supplementary data for representation of the other arrays. 

\begin{equation}\label{eq:RFE_vect_bcu}
\mathbf{S}^{\scaleto{RFE}{4pt}} = \begin{bmatrix}
0 &  s_{0}^{\scaleto{RFE}{4pt}} & s_{192}^{\scaleto{RFE}{4pt}} & s_{384}^{\scaleto{RFE}{4pt}} & 0\\
0 & s_{64}^{\scaleto{RFE}{4pt}} & s_{256}^{\scaleto{RFE}{4pt}} & s_{448}^{\scaleto{RFE}{4pt}} & s_{640}^{\scaleto{RBE}{4pt}}\\
0 & s_{32}^{\scaleto{RFE}{4pt}} & s_{224}^{\scaleto{RFE}{4pt}} & s_{416}^{\scaleto{RFE}{4pt}} & s_{608}^{\scaleto{RBE}{4pt}}\\
0 & s_{0}^{\scaleto{RFE}{4pt}}  & s_{192}^{\scaleto{RFE}{4pt}} & s_{384}^{\scaleto{RFE}{4pt}} & s_{576}^{\scaleto{RBE}{4pt}}\\
0 & 0 & 0 & 0 & 0\\
\end{bmatrix}
= \{\Vec{S}_0^{\scaleto{RFE}{4pt}}, \Vec{S}_1^{\scaleto{RFE}{4pt}}, \Vec{S}_2^{\scaleto{RFE}{4pt}}, \Vec{S}_3^{\scaleto{RFE}{4pt}}, \Vec{S}_4^{\scaleto{RFE}{4pt}}\}
\end{equation}

We also adopt a new notation, $\Vec{S}$, as shown in Eq. \ref{eq:RFE_vect_bcu}, with the same three-letter code related to the array in the superscript and a new subscript related to the column within the array. For example, $\Vec{S}_0^{\scaleto{RFE}{4pt}} = \mathbf{S}^{\scaleto{RFE}{4pt}}[0,:]$ and $\{\Vec{S}_0^{\scaleto{RFE}{4pt}}, \Vec{S}_1^{\scaleto{RFE}{4pt}}, \Vec{S}_2^{\scaleto{RFE}{4pt}}, \Vec{S}_3^{\scaleto{RFE}{4pt}}, \Vec{S}_4^{\scaleto{RFE}{4pt}}\}$ is a list of all the column vectors in $\mathbf{S}^{\scaleto{RFE}{4pt}}$.  We express the final array form as a list of column vectors as this is what is held in local memory of each PE.  For instance, 
%within five neighboring PEs on the WSE in this example (Fig \ref{fig:WSE_schematic}).  
all vectors with subscript 0, i.e. $\Vec{S}_0^{\scaleto{\mathrm{RFE}}{4pt}}$, $\Vec{S}_0^{\scaleto{\mathrm{RBO}}{4pt}}$, $\Vec{S}_0^{\scaleto{\mathrm{RBE}}{4pt}}$, $\Vec{S}_0^{\scaleto{\mathrm{RFO}}{4pt}}$,
$\Vec{S}_0^{\scaleto{\mathrm{BFE}}{4pt}}$,
$\Vec{S}_0^{\scaleto{\mathrm{BBO}}{4pt}}$,
$\Vec{S}_0^{\scaleto{\mathrm{BBE}}{4pt}}$, and
$\Vec{S}_0^{\scaleto{\mathrm{BFO}}{4pt}}$ 
are held on PE 0 ($k,i=0$), which is adjacent to PE 1 ($k,i=1$) that holds all vectors with subscript 1, and so forth (Figure \ref{fig:decomposition}b). In this notation, the PBC updates are executed as straightforward copy operations, as shown in Supplementary information.  

Algorithm \ref{alg:flipQuarter} outlines the implementation of the spin-flipping using the eight-array representation.  Initially, we employ Algorithm \ref{alg:pe_getNeighbors} to retrieve the neighbors of $\Vec{S}$. Subsequently, a set of half adders are applied to aggregate the bits into three $int16$ vectors, namely $\Vec{O}$, $\Vec{T}$, and $\Vec{F}$, representing the one's, two's, and four's place values of the neighbor spin sum.  Using a series of four masks, we obtain every fourth bit, enabling us to shift and combine  $\Vec{O}$, $\Vec{T}$ and  $\Vec{F}$ into a compacted four-bit integer ($\Vec{S}_{sum}[1:\m 1]$) which contains the sum of the neighbors of four spins in one $int16$ value. A masked $\Vec{S}$ is added to the eight's place to differentiate the sums as aligned or anti-aligned to the central spin. Subsequently, we loop over this final sum to extract each of the summed values as a single $int16$ (statement 13), which is then used to look up the acceptance ratio, $\Vec{A}_R$, from a precomputed exponent table.  A random float between 0 and 1, $\Vec{R}$, is generated and compared to $\Vec{A}_R$.  The outcome yields an $int16$ value of 0 or 1, which is shifted to the appropriate position, and then an exclusive OR is used to flip the bit, if the MC move is accepted.

We now have a complete set of kernels to implement the Ising model on the WSE using the massive parallelism available as shown in Algorithm \ref{alg:pe_program} (see also, supplementary document for PBE update algorithm).  While this implementation is more complex than those on traditional hardware, with this tailored approach to the WSE architecture, we obtain record acceleration of the Ising model.  

\section{Results and Discussion}\label{sec:results}
Thanks to Onsager\cite{Onsager1944PhysRev}, the 2D Ising model can be analytically solved for an infinite lattice size. Though computational methods are not strictly necessary for determining the phase transition, they are commonly utilized to assess the model's performance on new hardware and to gain insights for more intricate simulations such as 3D and higher Ising Hamiltonian's that are not amenable to the analytical solution. With this in mind, our aim is to create a straightforward 2D Ising model code for the WSE, compare its performance with other hardware, and affirm the WSE's viability for Monte Carlo-based modeling.

\subsection {Code Validation}

\begin{figure}[h!]
    \centering
\includegraphics[width=\textwidth]{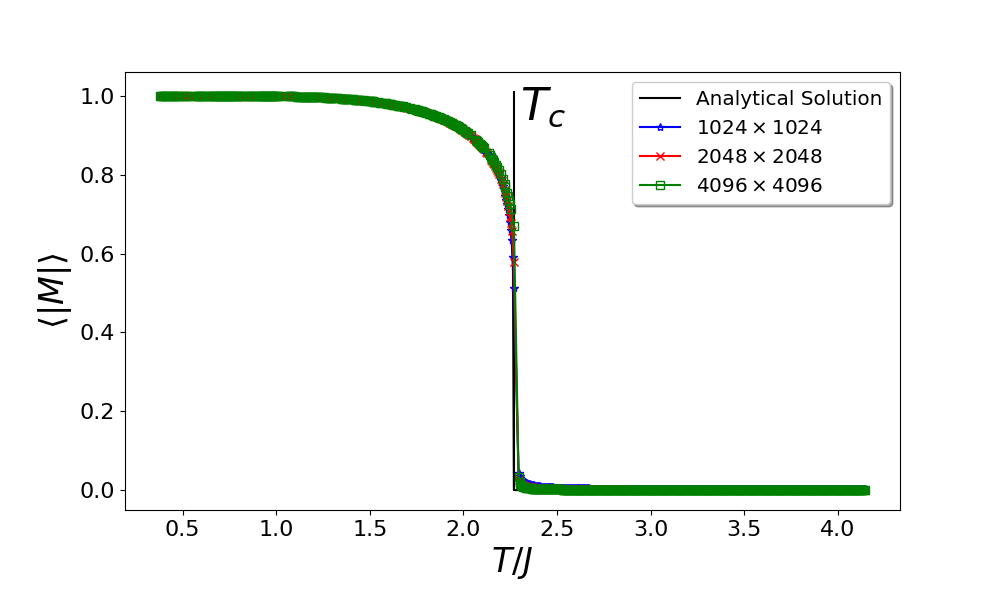}
    \caption{ The absolute average magnetization values obtained from Monte Carlo simulations on WSE for the 2D Ising model at temperatures between $0.385 J-4.145 J$ for three different sizes of 2D Ising lattices. For comparison, the analytical solution for the thermodynamic infinite large spin system size is also shown along with the critical temperature $T_c/J=2.269,185$ obtained from the analytic solution for infinite large spin system. 
    }
    \label{fig:M_valid}
\end{figure}

We first validate our implementation of the 2D Ising model on the WSE before we conduct a performance analysis.  Specifically,  we carried out 2262 interdependent simulations at three lattice sizes ($1024\times1024$, $2048\times2048$, and $4096\times4096$) for temperatures ranging from $T/J=0.385$ to $4.145$ in steps of $0.005$. For each simulation, we started from a completely random configuration of the Ising spins and carried out 12-24 million of MC iterations over the entire lattice. During the simulations, every 10,000-20,000 MC iterations, the mean value for the absolute magnetization value is calculated for post-analysis. Due to the large amount of data for post-analysis, an automatic method to discard the non-equilibration data  was used \cite{Chodera16JCTC}.  Figure~\ref{fig:M_valid}  shows the absolute average magnetization as a function of temperature. As seen from the figure, the average of the absolute magnetization values of the three lattice sizes matches the theoretical result predicted by Onsager where  $|M|={\left [ 1-\sinh^{-4}{(2J/T)} \right ]}^{1/8}$, and $|M|=0$ when $T \ge T_c$, and correctly predicted the phase transition temperature, $T_c/J=2.269,185$. These findings are strongly indicative of the validity of our implementation. We note here that we have ensured that the random number generator implemented in the MC simulation is of high fidelity, as detailed in the supplementary information.

\subsection {Performance Analysis} 
\begin{figure}[h!]
    \centering
    \includegraphics[width=\textwidth]{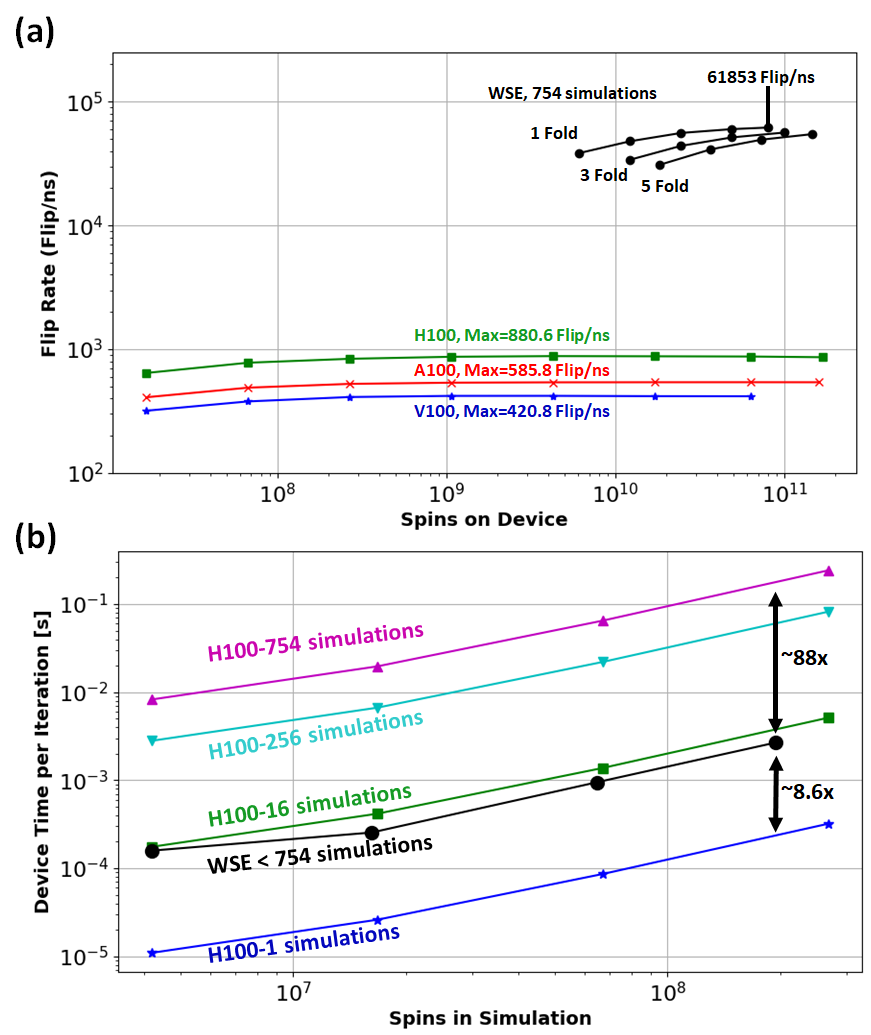}
    \caption{(a) The flip rate as a function of spins on device. The GPU V100 results are reported in \cite{Romero2020CPC} while as the A100 and H100 benchmarks are obtained in this study using the same code as in \cite{Romero2020CPC}. The WSE results are obtained using $N_{\rm{fold}}=1, 3,$ and 5 folding. 
    compares the productivity metric, defined as the cumulative device time per iteration to run $N_{\rm{sim}}$ simulations, between the H100 and WSE.
    (b) Productivity metric, defined as the  cumulative device time per iteration to run $N_{\rm{sim}}$ simulations. The lattice sizes for the WSE measurements are $2048 \times 2048$, $3952 \times 4096$, $7904 \times 8192$, and $11,586 \times 16,384$.  
    %The circles, triangles, and squares indicate 1, 3, and 5 folds in the PE axis respectively.  
    The lattice sizes for GPU measurements are $2048\times 2048$, $4096\times 4096$, $8192 \times 8192$, and $16,384\times 16,384$.}
    \label{fig:Total_Flips}
\end{figure}

Our initial performance evaluation primarily compares the net computation or flip rate $R_{\rm{flip}}$, measured in flip attempts per nanosecond (flip/ns). 
Figure \ref{fig:Total_Flips} summarizes our findings. The highest observed flip rate on a single WSE was $R_{\rm{flip}}=61,853$ flip/ns, achieved through our current compressed bit representation across 754 parallel simulations and with $N_{\rm{fold}}=1$. This marks a significant improvement over the performance reported by Romero et al. on V100\cite{Romero2020CPC}, which was over 142 times lower. Additionally, we conducted a comparison with modern GPUs by evaluating Romero's code on A100 and H100 GPUs, resulting in flip rates of 585.8 flips/ns and 880.6 flips/ns, respectively. Consequently, our WSE implementation demonstrates a flip rate advantage of over 70 times compared to the latest H100 GPUs.

In addition to the computational efficiency, another aspect worth discussing is the maximum lattice size feasible for simulation on the WSE. In this regard, 
%When comparing the capabilities between the GPU and our WSE implementation, it becomes apparent that 
the GPU can handle a significantly larger single lattice than what we implemented in the WFA for the WSE. Specifically, the largest lattice size we have managed on a single H100 or A100 with 80 GB memory is $409,600\times409,600$ totaling 168 billion spins. In contrast, our current code in the WFA allows with $N_{\rm{fold}}=5$ for a maximum single simulation of $11,856 \times 16,384$, comprising 194 million spins. While this difference might appear disadvantageous initially, and notwithstanding the existing analytical solution by Onsager, the practicality of running such extensive Ising simulations is limited due to the time required to achieve equilibrium in the MC simulation. Namely, although equilibrating the Ising system depends on factors such as temperature and initial configuration, it is reasonable to assume that the equilibration time scales with the total number of spins in the system. Thus, for a lattice size of $409,600^2$ spins and a 0.194-second iteration period obtained on H100, reaching equilibrium would necessitate over a thousand years of continuous computing -- an unfeasible task.

It is worth noting that the largest validation run conducted by Romero et al. involved a lattice of $4096\times4096$ spins.\cite{Romero2020CPC} For this size, each temperature required approximately 15 minutes on a V100 and 7.3 minutes on an H100 to achieve equilibrium. Considering these practical computing time constraints, the attainable size on a single row of PEs suffices. Therefore, we propose that conducting multiple simulations in parallel is more valuable than increasing the lattice size, given the significant time investment required for equilibrium in larger simulations.

Furthermore, as observed in Figure~\ref{fig:Total_Flips}, the GPU implementation\cite{Romero2020CPC} achieves its peak flip rate at smaller lattice sizes (ranging from $8192\times8192$ to $16,384\times16,384$) compared to the maximum lattice size accommodated in memory (up to $409,600\times409,600$ for H100), primarily due to the GPU fully utilizing its memory bandwidth. Consequently, any increase in the number of spins leads to a proportional increase in computation time per iteration. While the compressed bit representation\cite{Romero2020CPC} partly alleviates the memory bandwidth pressure by encoding more spins into each byte of data transferred between main memory and processors, reaching saturation at as few as $8192^2=67$ million spins implies limited advantages in developing methods for running more simulations in parallel on a single GPU. This limitation arises because there are no computing resources left to allocate to more than one simulation per device.

In contrast to traditional CPU or GPU hardware, the WSE is fundamentally different because all memory is accessible at L1 cache rates (128b read and 64b write per cycle on the current generation WSE).  In addition, 64b of data can be received/sent on the network-on-chip router per cycle.  As long as the data needed for computation exists in the local vicinity of a PE, WSE computations will be compute-bound, which is the case for our implementation of the 2D Ising model.  

To underscore this point, we present the simulation dimensions, flip rate $R_{\rm{flip}}$, and iteration period for $N_{\rm{fold}}=5$ benchmarks in Table~1. The iteration period $T_{\rm{iter}} = (n \times m \times N_{\rm{sim}})/R_{\rm{flip}}$ denotes the measured time it takes to attempt to flip all the spins on the device.  Notice that $T_{\rm{iter}}$ is independent of the lattice dimension assigned to one PE axis, $n$, and the number of independent simulations, $N_{\rm{sim}}$ assigned to the other PE axis, and only varies in response to the number of spins held in the memory of each PE (in proportion to $m$).  This is justified because the workload for each PE scales in direct proportion to the vectors that it must loop over, e.g., see  Eq. \ref{eq:RFE_vect_bcu}. The distinctive aspect of WSE computing lies in the fact that memory bandwidth and on-chip network bandwidth, scale directly with the number of PEs utilized in the simulation. This scalability ensures that the bandwidth is consistently adequate to support local operations proceeding at compute-bound rates. As a result, within this framework, neighbor-based algorithms such as the Ising model can achieve perfect weak scaling up to the edge of the wafer and very good strong scaling. This stands in stark contrast to traditional architectures where the main memory bandwidth is fixed and often significantly lower, by orders of magnitude, than what is required to sustain processors operating at compute-bound rates, as demonstrated by the performance on the GPUs shown in Figure~\ref{fig:Total_Flips}a. This results in poor strong scaling.

As a more meaningful metric, Figure~\ref{fig:Total_Flips}b compares the productivity metric, defined as the cumulative device time per iteration to run $N_{\rm{sim}}$ simulations, between the H100 and WSE. As seen from the figure, our implementation on the WSE can run up to 754 simulations in parallel so the device time per iteration is the same for a given lattice size up to 754 simulations.  However, the device time per iteration increases for the GPU in direct proportion to $N_{\rm{sim}}$.  This analysis shows that if more than $9$ simulations are needed, the WSE will be faster with our current code.  Further, at the limit of capacity (754 simulations), the WSE is as much as 88 times more productive than a H100.

It is worth noting that folding the spin array $N_{\rm{fold}}$ times and consolidating all resulting arrays into the same column of PEs does affect performance by increasing the workload per PE within each iteration. There are alternative mappings of this problem to the hardware that differ from the current implementation. For example, assigning each fold to its own column of PEs could reduce the PE workload by a factor of $N_{\rm{fold}} + 1$ and consequently decrease iteration time proportionally. Although this mapping would separate one PBC edge by $N_{\rm{fold}}$ PEs, it would only introduce $N_{\rm{fold}}-1$ cycles of latency to a PBC update. This increase is relatively minor, especially considering that PBC updates are infrequent. Therefore, as long as $N_{\rm{fold}}$ remains relatively small, its impact on computing speed is negligible. Additionally, this approach would enable larger individual simulation sizes, as more PEs (along with their associated memory) are allocated to each simulation. However, the productivity advantage depicted in Figure~\ref{fig:Total_Flips}b 
would remain largely unchanged, as the number of simulations running in parallel would decrease by $N_{\rm{fold}} + 1$. In essence, while the time per iteration for all parallel simulations would decrease by $N_{\rm{fold}} + 1$, the number of simulations running in parallel would also decrease by the same factor.
 
The 2D Ising model is positioned towards the lower end of the data intensity spectrum of HPC applications. Romero et al.\cite{Romero2020CPC} reported that they achieved 6 bits of IO per flip attempt in their highly optimized code, which is very near the theoretical minimum.  Keep in mind that many operations need to be performed per flip attempt, so the theoretical minimum data intensity is $0.5/N_{op}$ bytes per operation where $N_{op}$ is the number of operations in a flip attempt and is much greater than unity. It is useful to contrast the Ising model with a seven-point-stencil with non-constant coefficients in double precision, a very common HPC algorithm, where 104 bytes of IO are needed and 13 operations are performed per stencil evaluation giving it a data intensity of 8 bytes per operation.  Even before counting the operations in a particular Ising implementation, the data intensity of the Ising model is over an order of magnitude lower than this common HPC algorithm.  Given that low data intensity workloads favor traditional architectures due to the ability to hide computing behind low memory access rates with effective caching, and the fact that there is no ability to do this on the WSE, it is reasonable to infer that the current work will be amongst the lowest figure of merits reported in HPC algorithm ports.  Indeed, our previous work with computational fluid dynamics did find significantly higher figures of merits in the several hundred range.\cite{rocki2020Fast, woo2022disruptive}  In light of this, a figure of merit of 88 times greater productivity in comparison to the fastest available current-generation GPU is impressive and understandable relative to our previous work.

\section {Conclusions}

The Ising model's adaptability and broad applicability have solidified its position as a foundational tool in statistical physics and hardware assessment. Our study introduces a pioneering implementation of the two-dimensional Ising model on the Cerebras Wafer-Scale Engine (WSE), a revolutionary computing platform. We tailored the checkerboard algorithm for the unique WSE architecture, we used a compressed bit representation where 16 spins are represented as a single $int16$ work and performed the Monte Carlo simulations of the Ising model relying on computationally efficient bit manipulations. Importantly, to take advantage of the efficient communications between neighboring processing units on the WSE, we decomposed the Ising lattice into an eight array representation that resulted in a code that exhibits nearly perfect weak scaling as shown in Table 1.  Our implementation demonstrates impressive parallel performance, handling up to 754 simulations concurrently and achieving unprecedented flip attempt rates for Ising models with up to 200 million spins with nearly over 61.8 trillion flip attempts per second. This acceleration represents a gain of up to 148 times over previously reported single-device with a highly optimized implementation on NVIDIA V100 and up to 88 times in productivity compared to NVIDIA H100. Our findings highlight the significant potential of the WSE in scientific computing, particularly in the field of materials modeling.

\section {Declaration of competing interest}
The author declares no competing interests.

\section {Acknowledgments}
We acknowledge Michael James, Leighton Willson, and Kylee Santos at Cerebras for useful discussions about software half adders that enabled us to compute rapid sums and for discussions about scaling behavior and optimizations.  We would also like to thank Hal Finkel at Office of Science for access to the A100 GPUs on Perlmutter at National Energy Research Scientific Computing Center.  Finally, we would like to thank Prof. Sebastian Scherer from the Robotics Institute at Carnage Mellon University for facilitating access to NVIDIA H100 GPUs under the DURIP award W911NF2310067.

This project was funded by the U.S. Department of Energy, National Energy Technology Laboratory, in part, through a site support contract. Neither the United States Government nor any agency thereof, nor any of their employees, nor the support contractor, nor any of their employees, makes any warranty, express or implied, or assumes any legal liability or responsibility for the accuracy, completeness, or usefulness of any information, apparatus, product, or process disclosed, or represents that its use would not infringe privately owned rights. Reference herein to any specific commercial product, process, or service by trade name, trademark, manufacturer, or otherwise does not necessarily constitute or imply its endorsement, recommendation, or favoring by the United States Government or any agency thereof. The views and opinions of authors expressed herein do not necessarily state or reflect those of the United States Government or any agency thereof.

%\bibliographystyle{IEEEtran}
%\bibliography{IEEEabrv,main}
\bibliography{main}% common bib file

\newpage

\clearpage

\clearpage

\newpage
\clearpage

\newpage
\clearpage

\newpage
\clearpage

\begin{algorithm}
\caption{PE Level getNeighbors Pseudo-Code. The ``right'' operation refers to data transfer between neighboring PE ($k,i+1$) and center PE ($k,i$). While, the ``left'' operation refers to data transfer between neighboring PE ($k,i-1$) and center PE ($k,i$).   The $\gamma$ symbol is equivalent to $i$ to indicate the PE position in the $y$ direction. For example, in Eq. (\ref{eq:RFE_vect_bcu}), $\gamma$ varies between 1-- 3. The ``up'' and ``down'' refer to data transfer within a local memory, which resides in the current center PE ($k,i$).  The $|$ symbol indicates the bitwise OR operation. The $\gg$ and $\ll$ symbols indicate the bitwise right and left shift operations for the 16-bit strings.}\label{alg:pe_getNeighbors}
\begin{algorithmic}[1]
\Function{getNeighbors}{$\Vec{S}^a, \Vec{S}^b, \gamma, up, right$}
    \If{right}
        \State $\Vec{S}^R_{\gamma}[1:\m 1] \gets \Vec{S}_{\gamma+1}^a[1:\m1]$
        \State $\Vec{S}^L_{\gamma}[1:\m 1] \gets \Vec{S}_{\gamma}^a[1:\m 1]$
    \Else
        \State $\Vec{S}^R_{\gamma}[1:\m 1] \gets \Vec{S}_{\gamma}^a[1:\m1]$
        \State $\Vec{S}^L_{\gamma}[1:\m 1] \gets \Vec{S}_{\gamma-1}^a[1:\m1]$
    \EndIf
    \If{up}
        \State $\Vec{S}^T_{\gamma}[1:\m 1] = \Call{getBitAbove}{\Vec{S}_{\gamma}^b}$
        \State $\Vec{S}^B_{\gamma}[1:\m 1] \gets \Vec{S}_{\gamma}^b[1:\m1]$
    \Else
        \State $\Vec{S}^T_{\gamma}[1:\m 1] \gets \Vec{S}_{\gamma}^b[1:\m1]$
        \State $\Vec{S}^B_{\gamma}[1:\m 1] = \Call{getBitBelow}{\Vec{S}_{\gamma}^b}$
    \EndIf
    \State \Return{$\Vec{S}^R_{\gamma}, \Vec{S}^L_{\gamma}, \Vec{S}^T_{\gamma}, \Vec{S}^B_{\gamma}$}
\EndFunction
\Function{getBitAbove}{$\Vec{S}$}
    \State \Return{$(\Vec{S}[2:] \ll 15) \ | \ (\Vec{S}[1:\m 1] \gg 1)$}
\EndFunction
\Function{getBitBelow}{$\Vec{S}$}
    \State \Return{$(\Vec{S}[:\m 2] \gg 15) \ | \ (\Vec{S}[1:\m 1] \ll 1)$}
\EndFunction
\end{algorithmic}
\end{algorithm}

\begin{algorithm}
\caption{PE Level Pseudo-Code to flip the spins. The \ \&\ symbol indicates the bitwise AND operation. The $\oplus$ symbol indicates the bitwise exclusive OR operation.}\label{alg:flipQuarter}
\begin{algorithmic}[1]
\Function{flipQuarter}{$\Vec{S}, \Vec{S}^a, \Vec{S}^b, \gamma, up, right$}
    \State $\Vec{S}^R, \Vec{S}^L, \Vec{S}^T, \Vec{S}^B = getNeigbors(\Vec{S}^a, \Vec{S}^b, \gamma, up, right)$
    \State $\Vec{O}, \Vec{T}, \Vec{F} = \Call{bitwiseAdd}{\Vec{S}^R, \Vec{S}^L, \Vec{S}^T, \Vec{S}^B}$
    \State $mask := [0x1111, 0x2222, 0x4444, 0x8888]$
    \For{i in 0 to 3}
        \State $\Vec{S}_{sum}[1:\m1] \gets (\Vec{O}[1: \m 1] \ \&\ mask[i]) \gg i$
        \State $\Vec{T}_M[1:\m 1] \gets (\Vec{T}[1:\m 1] \ \&\ mask[i]) \gg i$
        \State $\Vec{F}_M[1:\m 1] \gets (\Vec{F}[1:\m 1] \ \&\ mask[i]) \gg i$
        \State $\Vec{S}_M[1:\m 1] \gets (\Vec{S}[1:\m 1] \ \&\ mask[i]) \gg i$
        \State $\Vec{S}_{sum}[1:\m 1] \gets \Vec{S}_{sum}[1:\m 1] + (\Vec{T}_M[1:\m1] \ll 1) + (\Vec{F}_M[1:\m 1] \ll 2)$
        \State $\Vec{S}_{sum}[1:\m 1] \gets \Vec{S}_{sum}[1:\m 1] + (\Vec{S}_M[1:\m1] \ll 3)$
        \For{ii in 0 to 3}
            \State $\Vec{A}_{R}[1:\m 1] \gets expTable[(\Vec{S}_{sum}[1:\m 1] \gg 4 \times ii) \ \&\ 15]$
            \State $\Vec{R}[1:\m 1] = \Call{randVect}$
            \State $\Vec{S}[1:\m1] \gets \Vec{S}[1:\m 1] \oplus ((\Vec{R}[1:\m1] < \Vec{A}_{R}[1:\m 1]) \ll (4 \times ii + i))$
        \EndFor
    \EndFor
\EndFunction
\Function{flipRed}{$\gamma$}
    \State \Call{flipQuarter}{$\Vec{S}^{\scaleto{RFE}{4pt}}, \Vec{S}^{\scaleto{BFE}{4pt}}, \Vec{S}^{\scaleto{BFO}{4pt}}, \gamma, F, F$}
    \State \Call{flipQuarter}{$\Vec{S}^{\scaleto{RBO}{4pt}}, \Vec{S}^{\scaleto{BBO}{4pt}}, \Vec{S}^{\scaleto{BBE}{4pt}}, \gamma, T, F$}
    \State \Call{flipQuarter}{$\Vec{S}^{\scaleto{RBE}{4pt}}, \Vec{S}^{\scaleto{BBE}{4pt}}, \Vec{S}^{\scaleto{BBO}{4pt}}, \gamma, F, T$}
    \State \Call{flipQuarter}{$\Vec{S}^{\scaleto{RFO}{4pt}}, \Vec{S}^{\scaleto{BFO}{4pt}}, \Vec{S}^{\scaleto{BFE}{4pt}}, \gamma, T, T$}
\EndFunction
\Function{flipBlue}{$\gamma$}
    \State \Call{flipQuarter}{$\Vec{S}^{\scaleto{BFE}{4pt}}, \Vec{S}^{\scaleto{RFE}{4pt}}, \Vec{S}^{\scaleto{RFO}{4pt}}, \gamma, F, T$}
    \State \Call{flipQuarter}{$\Vec{S}^{\scaleto{BBO}{4pt}}, \Vec{S}^{\scaleto{RBO}{4pt}}, \Vec{S}^{\scaleto{RBE}{4pt}}, \gamma, T, T$}
    \State \Call{flipQuarter}{$\Vec{S}^{\scaleto{BBE}{4pt}}, \Vec{S}^{\scaleto{RBE}{4pt}}, \Vec{S}^{\scaleto{RBO}{4pt}}, \gamma, F, F$}
    \State \Call{flipQuarter}{$\Vec{S}^{\scaleto{BFO}{4pt}}, \Vec{S}^{\scaleto{RFO}{4pt}}, \Vec{S}^{\scaleto{RFE}{4pt}}, \gamma, T, F$}
\EndFunction
\end{algorithmic}
\end{algorithm}

\begin{algorithm}
\caption{PE Level Multi-Spin Pseudo-Code}\label{alg:pe_program}
\begin{algorithmic}[1]
\Require $\myMatrix{T} \subset \mathbb{Z}^{2} \;|\; 0 \leq \gamma < N_{PE},\;0 \leq \beta < N_{sim} \; \forall (\gamma,\beta) \in \myMatrix{T}$
\While{not eqilibrated}
    \State \Call{flipRed}{$\gamma$} $\; \forall \;(\gamma, \beta) \in \myMatrix{T}$ \Comment{PE parallel}
    \State \Call{updateRedBC}{$\gamma$} $\; \forall \;(\gamma, \beta) \in \myMatrix{T}$ \Comment{PE parallel}
    \State \Call{flipBlue}{$\gamma$} $\; \forall \;(\gamma, \beta) \in \myMatrix{T}$ \Comment{PE parallel}
    \State \Call{updateBlueBC}{$\gamma$} $\; \forall \;(\gamma, \beta) \in \myMatrix{T}$ \Comment{PE parallel}
\EndWhile
\end{algorithmic}
\end{algorithm}

\clearpage

% \usepackage{booktabs}
\newpage
\begin{table}[]
\begin{tabular}{@{}ccccc@{}}
\caption{Benchmark results for WFA Ising model with five folds.  $n$ and $m$ are lattice dimensions for each simulation. $N_{\rm{sim}}$ is the number of $n \times m$ simulations run in parallel.  $n$ and $N_{\rm{sim}}$ determine the number of PEs applied to the block of simulations.  $m$ determines the length of memory vectors.\label{table:Five_fold_benchmarks}}\\

\toprule
$n$     & $m$     & $N_{\rm{sim}}$ & \begin{tabular}[c]{@{}c@{}}Flip Rate\\ (flip/ns)\end{tabular} & \begin{tabular}[c]{@{}c@{}}Iteration Period\\ (ms)\end{tabular} \\ \midrule
1536  & 2048  & 128    & 681                                                           & 0.591,48                                                        \\
3072  & 2048  & 256    & 2723                                                          & 0.591,56                                                         \\
6144  & 2048  & 512    & 10891                                                         & 0.591,56                                                         \\
11856 & 2048  & 754    & 30948                                                         & 0.591,57                                                        \\ \midrule
1536  & 4096  & 128    & 907                                                           & 0.888,24                                                         \\
3072  & 4096  & 256    & 3626                                                          & 0.888,31                                                         \\
6144  & 4096  & 512    & 14505                                                         & 0.888,32                                                         \\
11856 & 4096  & 754    & 41219                                                         & 0.888,32                                                         \\ \midrule
1536  & 8192  & 128    & 1085                                                          & 1.483,69                                                          \\
3072  & 8192  & 256    & 4342                                                          & 1.483,70                                                          \\
6144  & 8192  & 512    & 17368                                                         & 1.483,71                                                          \\
11856 & 8192  & 754    & 49357                                                         & 1.483,71                                                          \\ \midrule
1536  & 16384 & 128    & 1204                                                          & 2.675,25                                                          \\
3072  & 16384 & 256    & 4816                                                          & 2.675,25                                                          \\
6144  & 16384 & 512    & 19265                                                         & 2.675,24                                                          \\
11856 & 16384 & 754    & 54747                                                         & 2.675,26                                                          \\ \bottomrule
\end{tabular}
\end{table}
\clearpage

\section{Supplementary Information}

\subsection {Algorithms for Single-spin Multi-spin Updates}
\begin{algorithm}
\caption{2D single-spin  algorithm.}\label{alg:gen_ising}
\begin{algorithmic}[1]
\Require $\myMatrix{P}\subset \mathbb{Z}^{2} \;|\; 0 \leq i < N,\;0 \leq j < M \; \forall (i,j) \in \myMatrix{P}$
\State $\myMatrix{\sigma}[i,j] \gets random(\{\m1, 1\}) \; \forall (i,j) \in \myMatrix{P}$
\While{not eqilibrated}
    \State \Call{FlipSpin}{$i, j, \sigma$} $ \; \forall (i,j) \in \myMatrix{P}$ \Comment{any traversal order}
\EndWhile
\Function{FlipSpin}{i, j, $\sigma$}
    \State $\myMatrix{k} := \{(i+1, j), (i\m1, j), (i,j+1), (i,j\m1)\}$ \Comment{local adjacency matrix}
    \State $\myMatrix{k} = \Call{wrap}{\myMatrix{k}}$ \Comment{apply periodic boundary}
    \State $AR \gets exp(\m(2\myMatrix{\sigma}[i,j] \sum_{0 \leq \gamma < 3} \myMatrix{\sigma}[\myMatrix{k}[\gamma]])/T)$
    \If{$random(\{ $x$ \subset \mathbb{R}\;|\; 0<$x$ \leq 1\}) < AR$}
        \State $\myMatrix{\sigma}[i,j] \gets \m\myMatrix{\sigma}[i,j]$
    \EndIf
\EndFunction
\end{algorithmic}
\end{algorithm}

\begin{algorithm}
\caption{2D checkerboard multi-spin algorithm}\label{alg:ms_ising}
\begin{algorithmic}[1]
\Require $\myMatrix{P}\subset \mathbb{Z}^{2} \;|\; 0 \leq i < N,\;0 \leq j < M \; \forall (i,j) \in \myMatrix{P}$
\Require $\myMatrix{P}_{red} \subset \myMatrix{P} \;|\; mod(i+j,2)=0 \;\forall(i,j)\in \myMatrix{P}$
\Require $\myMatrix{P}_{blue} \subset \myMatrix{P} \;|\; mod(i+j,2)=1 \;\forall(i,j)\in \myMatrix{P}$
\State $\myMatrix{\sigma}[i,j] \gets random(\{\m 1, 1\}) \; \forall (i,j) \in \myMatrix{P}$
\While{not eqilibrated}
    \State \Call{FlipSpin}{$i, j, \sigma$} $ \; \forall (i,j) \in \myMatrix{P}_{red}$ \Comment{in parallel}
    \State \Call{FlipSpin}{$i, j, \sigma$} $ \; \forall (i,j) \in \myMatrix{P}_{blue}$ \Comment{in parallel}
\EndWhile
\Function{FlipSpin}{i, j, $\sigma$}
    \State $\myMatrix{k} := \{(i+1, j), (i \m 1, j), (i,j+1), (i,j\m1)\}$ \Comment{local adjacency matrix}
    \State $\myMatrix{k} = \Call{wrap}{\myMatrix{k}}$ \Comment{apply periodic boundary}
    \State $AR \gets exp(\m(2\myMatrix{\sigma}[i,j] \sum_{0 \leq \gamma < 3} \myMatrix{\sigma}[\myMatrix{k}[\gamma]])/T)$ \Comment{acceptance ratio calculation}
    \If{$random(\{ $x$ \subset \mathbb{R}\;|\; 0<$x$ \leq 1\}) < AR$}
        \State $\myMatrix{\sigma}[i,j] \gets -\myMatrix{\sigma}[i,j]$
    \EndIf
\EndFunction

\end{algorithmic}
\end{algorithm}

\newpage
\subsection {Algorithms for Periodic Boundary Condition Update}

\begin{algorithm}
\caption{PE Level BC Update Pseudo-Code}\label{alg:pe_bc_update}
\begin{algorithmic}[1]
\Function{updateBC}{$\Vec{S}, \Vec{S}^{nbr}, \gamma, up, right$}
    \If{up}
        \State $\Vec{S}_{\gamma}[\m 1] \gets \Vec{S}_{\gamma}[1]$
    \Else
        \State $\Vec{S}_{\gamma}[0] \gets \Vec{S}_{\gamma}[\m 2]$
    \EndIf
    \If{right}
        \If{$\gamma = N_{PE}$}
            \State $\Vec{S}_{\gamma}[1:\m 1] \gets \Vec{S}_{\gamma-1}^{nbr}[1:\m 1]$
        \ElsIf{$\gamma > 0$}
            \State $\Vec{T}_{\gamma}[1:\m 1] \gets \Vec{S}_{\gamma-1}^{nbr}[1:\m 1]$
        \EndIf
    \Else
        \If{$\gamma = 0$}
            \State $\Vec{S}_{\gamma}[1:\m 1] \gets \Vec{S}_{\gamma+1}^{nbr}[1:\m 1]$
        \ElsIf{$\gamma<N_{PE}$}
            \State $\Vec{T}_{\gamma}[1:\m 1] \gets \Vec{S}_{\gamma+1}^{nbr}[1:\m 1]$
        \EndIf
    \EndIf
\EndFunction
\Function{updateRedBC}{$\gamma$}
    \State \Call{updateBC}{$\Vec{S}^{\scaleto{RFE}{4pt}}, \Vec{S}^{\scaleto{RBE}{4pt}}, \gamma, T, T$}
    \State \Call{updateBC}{$\Vec{S}^{\scaleto{RBO}{4pt}}, \Vec{S}^{\scaleto{RFO}{4pt}}, \gamma, F, T$}
    \State \Call{updateBC}{$\Vec{S}^{\scaleto{RBE}{4pt}}, \Vec{S}^{\scaleto{RFE}{4pt}}, \gamma, T, F$}
    \State \Call{updateBC}{$\Vec{S}^{\scaleto{RFO}{4pt}}, \Vec{S}^{\scaleto{RBO}{4pt}}, \gamma, F, F$}
\EndFunction
\Function{updateBlueBC}{$\gamma$}
    \State \Call{updateBC}{$\Vec{S}^{\scaleto{BFE}{4pt}}, \Vec{S}^{\scaleto{BBE}{4pt}}, \gamma, T, F$}
    \State \Call{updateBC}{$\Vec{S}^{\scaleto{BBO}{4pt}}, \Vec{S}^{\scaleto{BFO}{4pt}}, \gamma, F, F$}
    \State \Call{updateBC}{$\Vec{S}^{\scaleto{BBE}{4pt}}, \Vec{S}^{\scaleto{BFE}{4pt}}, \gamma, T, T$}
    \State \Call{updateBC}{$\Vec{S}^{\scaleto{BFO}{4pt}}, \Vec{S}^{\scaleto{BBO}{4pt}}, \gamma, F, T$}
\EndFunction
\end{algorithmic}
\end{algorithm}

\newpage 
\subsection {12x96 Arrays After PBC Updates }

Eq. 2 in the main text references an example array out of a series of arrays after being expanded and PBC updated.  The following contains all arrays after expansion.  Eq. 1 in the main text references the arrays before expansion and PBC update.  Those arrays are simply the $3 \times 3$ central terms in the arrays below.

\begin{equation}\label{eq:RFE_vect_bcu}
\mathbf{S}^{\scaleto{RFE}{4pt}} = \begin{bmatrix}
0 &  s_{0}^{\scaleto{RFE}{4pt}} & s_{192}^{\scaleto{RFE}{4pt}} & s_{384}^{\scaleto{RFE}{4pt}} & 0\\
0 & s_{64}^{\scaleto{RFE}{4pt}} & s_{256}^{\scaleto{RFE}{4pt}} & s_{448}^{\scaleto{RFE}{4pt}} & s_{640}^{\scaleto{RBE}{4pt}}\\
0 & s_{32}^{\scaleto{RFE}{4pt}} & s_{224}^{\scaleto{RFE}{4pt}} & s_{416}^{\scaleto{RFE}{4pt}} & s_{608}^{\scaleto{RBE}{4pt}}\\
0 & s_{0}^{\scaleto{RFE}{4pt}}  & s_{192}^{\scaleto{RFE}{4pt}} & s_{384}^{\scaleto{RFE}{4pt}} & s_{576}^{\scaleto{RBE}{4pt}}\\
0 & 0 & 0 & 0 & 0\\
\end{bmatrix}
= \{\Vec{S}_0^{\scaleto{RFE}{4pt}}, \Vec{S}_1^{\scaleto{RFE}{4pt}}, \Vec{S}_2^{\scaleto{RFE}{4pt}}, \Vec{S}_3^{\scaleto{RFE}{4pt}}, \Vec{S}_4^{\scaleto{RFE}{4pt}}\}
\end{equation}

\begin{equation}\label{eq:RBO_vect_bcu}
\mathbf{S}^{\scaleto{RBO}{4pt}} = \begin{bmatrix}
0 & 0 & 0 & 0 & 0\\
0 & s_{1121}^{\scaleto{RBO}{4pt}} & s_{929}^{\scaleto{RBO}{4pt}} & s_{737}^{\scaleto{RBO}{4pt}} & s_{545}^{\scaleto{RFO}{4pt}}\\
0 & s_{1089}^{\scaleto{RBO}{4pt}} & s_{897}^{\scaleto{RBO}{4pt}} & s_{705}^{\scaleto{RBO}{4pt}} & s_{513}^{\scaleto{RFO}{4pt}}\\
0 & s_{1057}^{\scaleto{RBO}{4pt}} & s_{865}^{\scaleto{RBO}{4pt}} & s_{673}^{\scaleto{RBO}{4pt}} & s_{481}^{\scaleto{RFO}{4pt}}\\
0 & s_{1121}^{\scaleto{RBO}{4pt}} & s_{929}^{\scaleto{RBO}{4pt}} & s_{737}^{\scaleto{RBO}{4pt}} & 0\\
\end{bmatrix}
= \{\Vec{S}_0^{\scaleto{RBO}{4pt}}, \Vec{S}_1^{\scaleto{RBO}{4pt}}, \Vec{S}_2^{\scaleto{RBO}{4pt}}, \Vec{S}_3^{\scaleto{RBO}{4pt}}, \Vec{S}_4^{\scaleto{RBO}{4pt}}\}
\end{equation}

\begin{equation}\label{eq:RBE_vect_bcu}
\mathbf{S}^{\scaleto{RBE}{4pt}} = \begin{bmatrix}
0 &                            s_{960}^{\scaleto{RBE}{4pt}} & s_{768}^{\scaleto{RBE}{4pt}} & s_{576}^{\scaleto{RBE}{4pt}} & 0\\
s_{64}^{\scaleto{RFE}{4pt}} & s_{1024}^{\scaleto{RBE}{4pt}} & s_{832}^{\scaleto{RBE}{4pt}} & s_{640}^{\scaleto{RBE}{4pt}} & 0\\
s_{32}^{\scaleto{RFE}{4pt}} &  s_{992}^{\scaleto{RBE}{4pt}} & s_{800}^{\scaleto{RBE}{4pt}} & s_{608}^{\scaleto{RBE}{4pt}} & 0\\
 s_{0}^{\scaleto{RFE}{4pt}} &  s_{960}^{\scaleto{RBE}{4pt}} & s_{768}^{\scaleto{RBE}{4pt}} & s_{576}^{\scaleto{RBE}{4pt}} & 0\\
0 & 0 & 0 & 0 & 0\\
\end{bmatrix}
= \{\Vec{S}_0^{\scaleto{RBE}{4pt}}, \Vec{S}_1^{\scaleto{RBE}{4pt}}, \Vec{S}_2^{\scaleto{RBE}{4pt}}, \Vec{S}_3^{\scaleto{RBE}{4pt}}, \Vec{S}_4^{\scaleto{RBE}{4pt}}\}
\end{equation}

\begin{equation}\label{eq:RFO_vect_bcu}
\mathbf{S}^{\scaleto{RFO}{4pt}} = \begin{bmatrix}
0 & 0 & 0 & 0 & 0\\
s_{1121}^{\scaleto{RBO}{4pt}} & s_{161}^{\scaleto{RFO}{4pt}} & s_{353}^{\scaleto{RFO}{4pt}} & s_{545}^{\scaleto{RFO}{4pt}} & 0\\
s_{1089}^{\scaleto{RBO}{4pt}} & s_{129}^{\scaleto{RFO}{4pt}} & s_{321}^{\scaleto{RFO}{4pt}} & s_{513}^{\scaleto{RFO}{4pt}} & 0\\
s_{1057}^{\scaleto{RBO}{4pt}} &  s_{97}^{\scaleto{RFO}{4pt}} & s_{289}^{\scaleto{RFO}{4pt}} & s_{481}^{\scaleto{RFO}{4pt}} & 0\\
0                             &  s_{161}^{\scaleto{RFO}{4pt}} & s_{353}^{\scaleto{RFO}{4pt}} & s_{545}^{\scaleto{RFO}{4pt}} & 0\\
\end{bmatrix}
= \{\Vec{S}_0^{\scaleto{RFO}{4pt}}, \Vec{S}_1^{\scaleto{RFO}{4pt}}, \Vec{S}_2^{\scaleto{RFO}{4pt}}, \Vec{S}_3^{\scaleto{RFO}{4pt}}, \Vec{S}_4^{\scaleto{RFO}{4pt}}\}
\end{equation}

\begin{equation}\label{eq:BFE_vect}
\mathbf{S}^{\scaleto{BFE}{4pt}} = \begin{bmatrix}
0                             &  s_{96}^{\scaleto{BFE}{4pt}} & s_{288}^{\scaleto{BFE}{4pt}} & s_{480}^{\scaleto{BFE}{4pt}} & 0\\
s_{1120}^{\scaleto{BBE}{4pt}} & s_{160}^{\scaleto{BFE}{4pt}} & s_{352}^{\scaleto{BFE}{4pt}} & s_{544}^{\scaleto{BFE}{4pt}} & 0\\
s_{1088}^{\scaleto{BBE}{4pt}} & s_{128}^{\scaleto{BFE}{4pt}} & s_{320}^{\scaleto{BFE}{4pt}} & s_{512}^{\scaleto{BFE}{4pt}} & 0\\
s_{1056}^{\scaleto{BBE}{4pt}} &  s_{96}^{\scaleto{BFE}{4pt}} & s_{288}^{\scaleto{BFE}{4pt}} & s_{480}^{\scaleto{BFE}{4pt}} & 0\\
0 & 0 & 0 & 0 & 0\\
\end{bmatrix}
= \{\Vec{S}_0^{\scaleto{BFE}{4pt}}, \Vec{S}_1^{\scaleto{BFE}{4pt}}, \Vec{S}_2^{\scaleto{BFE}{4pt}}, \Vec{S}_3^{\scaleto{BFE}{4pt}}, \Vec{S}_4^{\scaleto{BFE}{4pt}}\}
\end{equation}

\begin{equation}\label{eq:BBO_vect_bcu}
\mathbf{S}^{\scaleto{BBO}{4pt}} = \begin{bmatrix}
0 & 0 & 0 & 0 & 0\\
s_{065}^{\scaleto{BFO}{4pt}} & s_{1025}^{\scaleto{BBO}{4pt}} & s_{833}^{\scaleto{BBO}{4pt}} & s_{641}^{\scaleto{BBO}{4pt}} & 0\\
s_{033}^{\scaleto{BFO}{4pt}} &  s_{993}^{\scaleto{BBO}{4pt}} & s_{801}^{\scaleto{BBO}{4pt}} & s_{609}^{\scaleto{BBO}{4pt}} & 0\\
s_{001}^{\scaleto{BFO}{4pt}} &  s_{961}^{\scaleto{BBO}{4pt}} & s_{769}^{\scaleto{BBO}{4pt}} & s_{577}^{\scaleto{BBO}{4pt}} & 0\\
0                            & s_{1025}^{\scaleto{BBO}{4pt}} & s_{833}^{\scaleto{BBO}{4pt}} & s_{641}^{\scaleto{BBO}{4pt}} & 0\\
\end{bmatrix}
= \{\Vec{S}_0^{\scaleto{BBO}{4pt}}, \Vec{S}_1^{\scaleto{BBO}{4pt}}, \Vec{S}_2^{\scaleto{BBO}{4pt}}, \Vec{S}_3^{\scaleto{BBO}{4pt}}, \Vec{S}_4^{\scaleto{BBO}{4pt}}\}
\end{equation}

\begin{equation}\label{eq:BBE_vect_bcu}
\mathbf{S}^{\scaleto{BBE}{4pt}} = \begin{bmatrix}
0 & s_{1056}^{\scaleto{BBE}{4pt}} & s_{864}^{\scaleto{BBE}{4pt}} & s_{672}^{\scaleto{BBE}{4pt}} & 0\\
0 & s_{1120}^{\scaleto{BBE}{4pt}} & s_{928}^{\scaleto{BBE}{4pt}} & s_{736}^{\scaleto{BBE}{4pt}} & s_{544}^{\scaleto{BFE}{4pt}}\\
0 & s_{1088}^{\scaleto{BBE}{4pt}} & s_{896}^{\scaleto{BBE}{4pt}} & s_{704}^{\scaleto{BBE}{4pt}} & s_{512}^{\scaleto{BFE}{4pt}}\\
0 & s_{1056}^{\scaleto{BBE}{4pt}} & s_{864}^{\scaleto{BBE}{4pt}} & s_{672}^{\scaleto{BBE}{4pt}} & s_{480}^{\scaleto{BFE}{4pt}}\\
0 & 0 & 0 & 0 & 0\\
\end{bmatrix}
= \{\Vec{S}_0^{\scaleto{BBE}{4pt}}, \Vec{S}_1^{\scaleto{BBE}{4pt}}, \Vec{S}_2^{\scaleto{BBE}{4pt}}, \Vec{S}_3^{\scaleto{BBE}{4pt}}, \Vec{S}_4^{\scaleto{BBE}{4pt}}\}
\end{equation}

\begin{equation}\label{eq:BFO_vect_bcu}
\mathbf{S}^{\scaleto{BFO}{4pt}} = \begin{bmatrix}
0 & 0 & 0 & 0 & 0\\
0 & s_{065}^{\scaleto{BFO}{4pt}} & s_{257}^{\scaleto{BFO}{4pt}} & s_{449}^{\scaleto{BFO}{4pt}} & s_{641}^{\scaleto{BBO}{4pt}}\\
0 & s_{033}^{\scaleto{BFO}{4pt}} & s_{225}^{\scaleto{BFO}{4pt}} & s_{417}^{\scaleto{BFO}{4pt}} & s_{609}^{\scaleto{BBO}{4pt}}\\
0 & s_{001}^{\scaleto{BFO}{4pt}} & s_{193}^{\scaleto{BFO}{4pt}} & s_{385}^{\scaleto{BFO}{4pt}} & s_{577}^{\scaleto{BBO}{4pt}}\\
0 & s_{065}^{\scaleto{BFO}{4pt}} & s_{257}^{\scaleto{BFO}{4pt}} & s_{449}^{\scaleto{BFO}{4pt}} & 0\\
\end{bmatrix}
= \{\Vec{S}_0^{\scaleto{BFO}{4pt}}, \Vec{S}_1^{\scaleto{BFO}{4pt}}, \Vec{S}_2^{\scaleto{BFO}{4pt}}, \Vec{S}_3^{\scaleto{BFO}{4pt}}, \Vec{S}_4^{\scaleto{BFO}{4pt}}\}
\end{equation}

\newpage
\subsection {Pseudo-Random Bitstream Tests}

The WSE has a Pseudo-Random Number Generator (PRNG) that is capable of generating 64 bits of pseudo random data per cycle on each PE.  To start our investigation into the quality of the PRNG, we wrote a bit perfect representation of the PRNG in Python (not available due to a NDA) which allows us to study the PRNG appart from hardware.  At its heart, the PRNG generates a random stream of bits which can later be shaped into a particular set of numbers according to a binary format.  Several methods have been developed to assess the quality of of PRNG bit stream.  We chose to test according to NIST SP 800-22 Rev. 1.\cite{NistRandomStandard} This standard was developed to assess PRNGs for suitability in stringent cryptographic applications where approximation of randomness are exceptionally important.  To quote Ruken et. al. "Generators suitable for use in cryptographic applications may need to meet stronger requirements than for other applications. In particular, their outputs must be unpredictable in the absence of knowledge of the inputs." \cite{NistRandomStandard}  The rigorous statistical tests in NIST SP 800-22 Rev. 1 are intended to assess any stream of bits for uniformity, independence, and unpredictability.  The tests calculate the probability of a bit stream having the character of a truly random sequence and compare it to a defined threshold specified for each test to determine if bit sequence is "random" or "non-random".  Most of the tests recommend a sequence of bits larger than 1,000,000 bits, so we generated a sequence of 6,400,000 bits with the Python PRNG code.  There are several open-source libraries available to run the statistical tests given a bit sequence and we chose the "NIST Randomness Test Suit".\cite{NistRandomnessTestSuite} 

The bit stream was random according to the Frequency (Monobit) Test, Frequency Test within a Block Test, Runs Test, Longest Run of Ones in a Block Test, Discrete Fourier Transform Test, Non-overlapping Template Matching Test, Overlapping Template Matching Test, Maurer's "Universal Statistical" Test, Linear Complexity Test, Serial Test, Approximate Entropy Test, Cumulative Sums Test, Random Excursions Test, and Random Excursions Variant Test.  From these tests, we can conclude that the WSE is capable of generating high quality pseudo-random bit streams having the appearance of uniformity, independence, and unpredictability.

\subsection {Pseudo-Random Ranged Float Tests}
A random stream of bits needs to be shaped to produce meaningful float values as a random bit arrangement could produce any float value in the full dynamic range as well as values that cannot be interpreted as a float according to the IEEE 754 standard.  In particular, we are interested in generating random values in the range of 0.0 and 1.0.  To do this, the most significant 10 bits are discarded via an AND with 0x7fffff.  The result is then combined with 0x3f800000 with an OR to create a pseudo random float in the range of 1.0 to 2.0.  1.0 is then subtracted to yield a pseudo random float in the range of 0.0 to 1.0.  Given the rigorous tests above, it is highly likely that the resulting ranged floats also have random character.  To verify this, we did further examinations for uniformity, independence and unpredictability.

\begin{figure}[h!]
    \centering
\includegraphics[width=1.0 \textwidth]{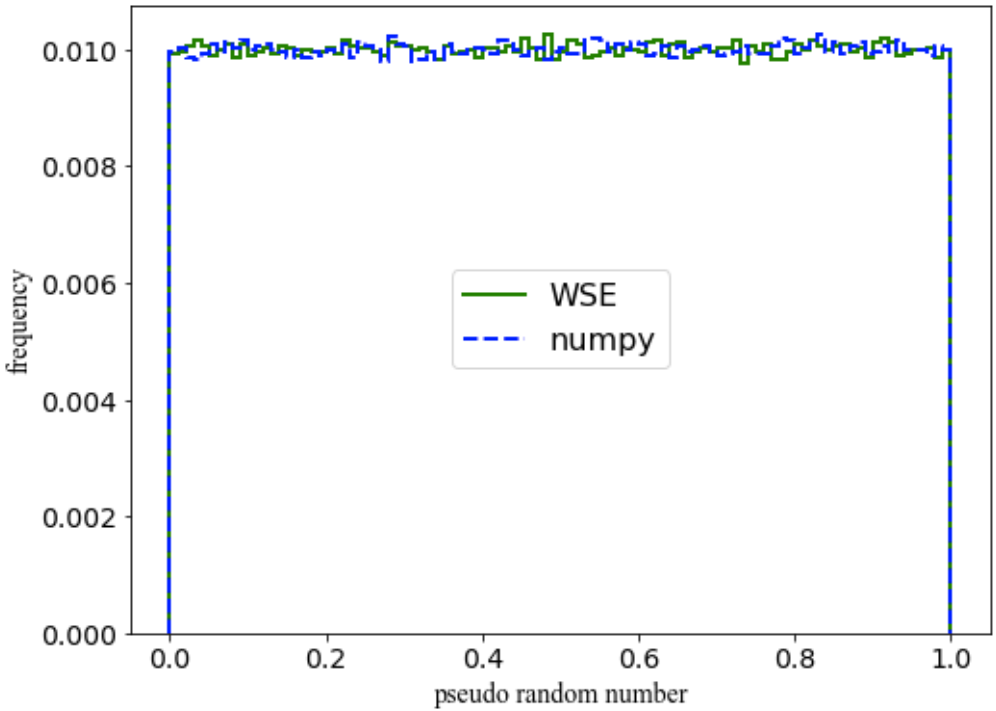}
    \caption{The frequency analysis for two sets of pseudo-random numbers with a size of 1000000. a)The pseudo random numbers generated on WSE. b) The pseudo random numbers generated from numpy by using the numpy.random.rand. 
    }
\label{fig:WSE_numpy_frequency}
\end{figure}

We applied the transformations above to generate random floats between 0 and 1 and confirmed its accuracy against WSE hardware.  We generated a random number array with shape of (2,2,1000000), which is denoted as WSE-rand(2,2,1000000). To evaluate the uniformity of the random numbers, the frequencies were calculated for 1 million random numbers in 100 blocks of [0, 0.01), [0.01, 0.02), [0.02, 0.03), ..[0.99, 1.0) with a bin width of 0.01. All the values in the frequency data array were found to be close to 0.01, as expected. The highest frequency value is 0.010422, and the lowest frequency value is 0.009755. The relative difference between the highest and the lowest frequency was found to be 6.8\% (see Fig. \ref{fig:WSE_numpy_frequency}).  When a larger bin width of 0.1 was chosen, the relative difference between the highest and the lowest frequency is decreased to 1.1\%.  

\begin{figure}[h!]
    \centering
\includegraphics[width=\textwidth]{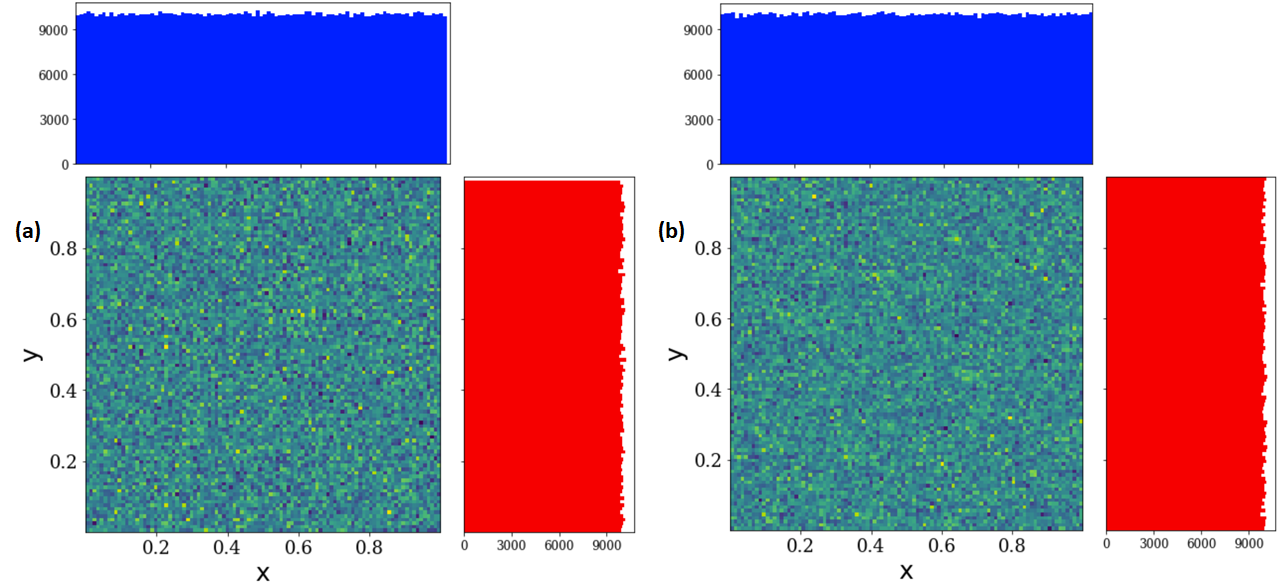}
    \caption{The 2-dimensional histogram analysis for two variables $x$ and $y$. Also shown are the 1 dimensional histogram analysis for $x$ (top) and $y$ (right) respectively. For a random sequence array, such as rand[0:1000000], $x=$rand[0:-1], and $y=$rand[1:]. The a) is for the pseudo random numbers generated on WSE, and b) is for the pseudo random numbers  generated from numpy by using numpy.random.rand. 
    }
\label{fig:WSE_rand_joint}
\end{figure}

To evaluate the  independence of the random numbers, the joint distribution for two variables are calculated. The two random variables are set to be $x=$WSE-rand[i,j,0:-1] and $y=$WSE-rand[i,j,1:] (i,j=0,1). One example of the joint distribution between $x$ and $y$ for the WSE-rand(1,1,:) is shown in Fig. \ref{fig:WSE_rand_joint}. There is no obvious pattern in Fig.\ref{fig:WSE_rand_joint} and the axis histograms are flat and uniform, which suggests that $x$ and $y$ are independent from each other. The other three random number sequences, WSE-rand(0,0,:), WSE-rand(0,1,:) and WSE-rand(1,0,:) show similar independence as Fig.\ref{fig:WSE_rand_joint}.  

For comparison, we also calculated uniform, independence, and unpredictability for random numbers generated by using Numpy (ie: np.random.rand(2,2,1000000)). Similarly to the random numbers generated on WSE, all the frequency values are close to 0.01 when the bin width was chosen to be 0.01. The highest frequency value is 0.010271, and the lowest frequency value is 0.009708. The relative difference between the highest and the lowest frequency is 5.8\% (Fig. \ref{fig:WSE_numpy_frequency}). When the bin width is increased to 0.1, the relative difference between the highest and the lowest frequency is decreased to be 1.2\%. The independence of the random number of was also evaluated and $x$ and $y$ values for the np.random.rand show the similar behavior as Fig.\ref{fig:WSE_rand_joint}. The simple Runs Test also suggest that the numbers are unpredictable and random. 

As a final test, we generated 8,388,608 ($2^{23}$) random floats and converted them to a bit sequence by an element wise comparison to the median value of the sequence (bit[i] = float[i] $\ge$ median(float[:]).  The bit sequence was fed into The NIST Random Test Suite as before.  If a PRNG is used to generate a set of random values within a range, each random value should have an equal probability of being above or below the median.  Thus the bit sequence in this test should also pass the random character tests in the NIST SP 800-22 Rev. 1 standard.

The bit stream was random according to the Frequency (Monobit) Test, Frequency Test within a Block Test, Runs Test, Longest Run of Ones in a Block Test, Discrete Fourier Transform Test, Non-overlapping Template Matching Test, Overlapping Template Matching Test, Maurer's "Universal Statistical" Test, Serial Test, Approximate Entropy Test, Cumulative Sums Test, Random Excursions Test, and Random Excursions Variant Test.  From these tests, we can conclude that the methods used to transform the bit stream to a ranged set of psudo-random floats is also of high quality having the appearance of uniformity, independence, and unpredictability.

\newpage
\clearpage

\newpage
\clearpage

%\end{document}

\end{document}